\documentclass[aps,prb,twocolumn,superscriptaddress,showpacs]{revtex4-1}
\pdfoutput=1

\usepackage{epstopdf}
\usepackage{epsfig}
\usepackage{amsmath}
\usepackage{amssymb}
\usepackage[colorlinks=true,citecolor=blue,linkcolor=blue]{hyperref}
\newcommand{\ket}[1]{\left|#1\right>}

\newcommand{\braket}[1]{\left<#1\right>}
\newcommand{\para}[1]{\left(#1\right)}
\newcommand{\abs}[1]{\left|#1\right|}
\newcommand{\ld}[0]{\ \ \ \ \ \ }
\newcommand{\sd}[0]{\ \ \ }

\begin{document}

\title{Phase diagram of the strongly correlated Kane-Mele-Hubbard model}

\author{Abolhassan Vaezi}
\email{Corresponding author: vaezi@ipm.ir}
\affiliation{School of Physics, Institute for Research in Fundamental Sciences, IPM, Tehran, 19395-5531, Iran}
\affiliation{Department of Physics, Sharif University of Technology, Tehran 11155-9161, Iran}

\author{Mahdi Mashkoori}
\affiliation{Department of Physics, Sharif University of Technology, Tehran 11155-9161, Iran}

\author{Mehdi Hosseini}
\affiliation{Department of Physics, Sharif University of Technology, Tehran 11155-9161, Iran}
\affiliation{Department of Physics, Shiraz University of Technology, Shiraz 313-71555, Iran}

\begin{abstract}
 We explore the phase diagram of the strongly correlated Hubbard model with intrinsic spin orbit coupling on the honeycomb lattice. We obtain the low energy effective model describing the spin degree of freedom. We study the resulting model within the Schwinger boson and Schwinger fermion approaches. The Schwinger boson approach gives the boundary between the spin liquid phase and the magnetically ordered phases, Neel order and incommensurate Neel order. We find that increasing the strength of the spin orbit coupling, narrows the width of the spin liquid region. The Schwinger fermion approach sheds further light on the nature of the spin liquid phase. We obtain three different candidates for the spin liquid phase within the mean field approximation which are gapless spin liquid, topological Mott insulator, and the chiral spin liquid phases. We argue that the gauge fluctuations and the instanton effect may suppress the first two spin liquids, while the chiral spin liquid is stable against gauge fluctuations due to its nontrivial topology.
\end{abstract}
\date{\today}

%\pacs{71.10.Fd,71.10.Pm,73.20.-r}

%71.10.Fd   Lattice fermion models (Hubbard model, etc.)
%72.80.Ga Transition-metal compounds, electrical conductivity of
% 03.65.Vf Topological phases (quantum mechanics)
%71.10.Pm Fermions in reduced dimensions (anyons, composite fermions, Luttinger liquid, etc.)
%73.20.-r   Electron states at surfaces and interfaces

\maketitle
\section{Introduction}

The spin-orbit interaction driven topological insulator(TI) phase of matter with gapped bulk and protected gapless edge excitations has spurred a renewed interest in the topological states of matter. Kane and Mele proposed the first and the simplest model exhibiting TI phase for noninteracting electrons on the two dimensional honeycomb lattice \cite{Kane_Mele_2005_1,Kane_Mele_2005_2}. The topological nature of the TI phase and its $Z_2$ structure due to the time reversal invariance (TRI), protects the edge excitation from Anderson localization \cite{Zhang_Qi_2010_a,Hasan_2010_1}. The absence of localization means these systems should be experimentally realizable as they are \cite{Bernevig_2006_1,Hsieh_2008_1}. One crucial question that naturally arises is what happens to the TI starting from interacting electron systems. The simplest interaction term is the onsite Hubbard repulsion that causes strong correlation between electrons. Recently, Rachel and Le Hur in Ref. \cite{LeHur_2010_a} have studied the phase diagram of the Kane-Mele model in the presence of onsite Hubbard interaction. They concluded that the TI phase is stable against onsite repulsion up to a critical value of the interaction strength. Their result has been verified by several other authors both numerically and analytically \cite{Assad_2010_a,Wu_2010_a,LeHur_2011_a,Varney_2011_a,Fiete_2011_b,vaezi_2011_d,Yu_2011_a,Yamaji_2010_a}.

The phase diagram of the Hubbard model has been extensively studied and various techniques support the existence of a spin liquid phase proximate to the Mott metal-insulator transition point \cite{Meng_2010a,Vaezi_2010a,Mosadeq_2010_a,Clark_2010_a,Reuther_2011_a,Albuquerque_2011_a,Tran_2011_a,Ying_2010_1,Wang_Fa_2010a,Cenke_1}. In this paper, we are going to address the issue of the possibility of observing the spin liquid phase in the phase diagram of the Kane-Mele-Hubbard model. Spin liquids may occur when the charge degree of freedom is gapped and as a result frozen due to the interaction (Mott insulator), though the spin degree of freedom can be either gapless or gapped. Consequently, we can integrate out the charge degree of freedom as it is gapped with a gap of the order of onsite repulsion in the strong correlation regime. The resulting model describes the spin degree of freedom only. In the pure Hubbard model, that procedure leads to the derivation of the extended $J1-J2$ Heisenberg model. Adding spin-orbit interaction to the Hubbard model, through the Kane-Mele term, introduces a new term for the second neighbor that is of the form $g_{2} \para{-{\rm S}_x(i){\rm S}_x(j)-{\rm S}_y(i){\rm S}_y(j)+{\rm S}_z(i){\rm S}_z(j)}$, where $g_2=4\lambda_{\rm SO}^2/U$ \cite{LeHur_2010_a}. Combining this term with the Heisenberg interaction for the next nearest neighbor (NNN) yields an anisotropic XXZ model for the second nearest neighbor, while the nearest neighbor Heisenberg interaction remains intact. In this paper, we study the rich phase diagram of this model. Using a combination of the mean field results, gauge theory, instanton effect and topological arguments, we demonstrate that the phase diagram hosts a region of chiral gapped spin liquid phase \cite{He_2011_a} up to a critical value of the $g_2/J_2$ and for large enough $J_2/J_1$. An interesting possibility is the emergence of the gapped topological spin liquid phase (topological Mott insulator) which is the same as TI phase, except that its charge degree of freedom is gapped. For large enough values of the $g_2/J_2$, we show that the in plane XY magnetic ordering wins over the topological spin liquids after taking instanton effect into consideration. For small values of $J_2/J_1$ and $g_2/J_1$, we obtain a gapless spin liquid which is shown to be unstable toward Neel order or valence bond solid (VBS) state.

This paper is organized as follows: in section II, we introduce our model that describes the effective action for the spin degree of freedom in the strongly correlated Kane-Mele-Hubbard model and we derive Kane-Mele-Heisenberg (KMH) model. Section III aims at studying KMH model using Schwinger boson approach. Within the mean field approximation, we investigate the possibility of the spin liquid phase in KMH model. Our study yields a phase diagram with 1- Neel order 2- incommensurate Neel order 3- gapped spin liquid phase within the mean field level. Then we focus on the spin liquid phase and study it further in section IV using the Schwinger fermion approach. Within the mean field approximation we study the competition between three different kinds of spin liquids. In section V, we discuss the gauge theory of the Schwinger fermion mean field states. We argue that the microscopic KMH enjoys an SU(2) gauge degree of freedom. However the SU(2) group may break down to its U(1) or $Z_2$ subgroup in the mean field state after Anderson-Higgz mechanism. Moreover, a mean field state with nonzero Hall conductance can suppress gauge fluctuations further. We also take the effect of instantons into consideration to have a more accurate accounting of gauge fluctuations. Using topological arguments and gauge theory, in section VI, we argue that among the proposed spin liquids, the chiral gapped spin liquid with nonzero Hall conductance for Schwinger fermions is stable against gauge fluctuations, while two other states may undergo transition to spontaneously broken symmetry phases.

\section{Model}
The Kane-Mele-Hubbard model on the honeycomb lattice is described as
\begin{equation}
\label{KMH-model}
H=-t \sum_{\langle ij \rangle,\sigma} c^{\dag}_{i,\sigma}c_{j,\sigma}+U\sum_{i} n_{i\uparrow}n_{i\downarrow} + i\lambda_{SO}\sum_{\langle\langle i,j \rangle\rangle,\sigma}\sigma\nu_{ij}c^{\dag}_{i,\sigma}c_{j,\sigma}
\end{equation}
where $t$, $U$, and $\lambda_{\rm{SO}}$ are the nearest neighbor hopping energy, the strength of the on-site repulsion, and  the second-neighbor spin-orbit coupling strength, respectively. Here $c_{i\sigma}$ ($c_{i\sigma}^{\dag}$) annihilates (creates) an electron with spin $\sigma$ on site $i$. $\nu_{i,j}$ is introduced so as to obtain a nonzero flux turning around any triangular path and is defined as $\nu_{i,j}=\frac{\vec{d}_{i}\times \vec{d}_{j}}{\abs{\vec{d}_{i}\times \vec{d}_{j}}}. \hat{z}$ where $d_{i}$ and $d_j$ are two shortest vectors that connect sites $i$ and $j$ i.e. $\vec{d}_i+\vec{d}_j=\vec{R}_i-\vec{R}_j$ (see Fig. [1]). Since we are interested in studying the interplay between the strong correlation and the topology of the band structure, we only consider the intermediate and the large $U/t$ and $U/\lambda_{\rm{SO}}$ limit of the Kane-Mele-Hubbard model.

\begin{figure}[tbp]
\begin{center}
\includegraphics[width=120pt]{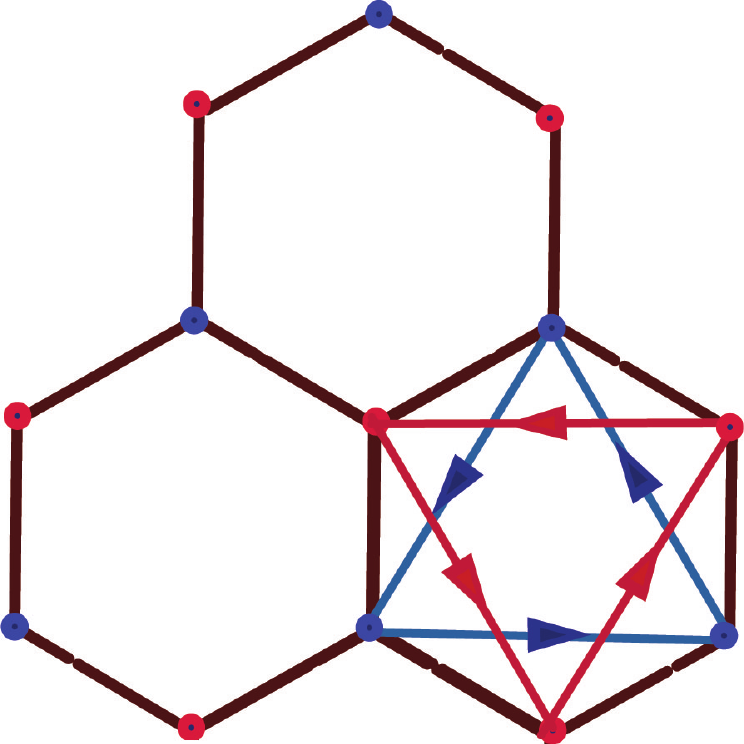}
\caption{(Color online) Kane-Mele model for the topological insulators. Arrows denote the phase of the hopping to the NNN for the spin up electrons. The phase for the spin downs is opposite to that of spin up electrons. Assigning a phase $\phi_{\sigma}$ to the hopping of spin $\sigma$ electrons induces a $3\phi_{\sigma}$ magnetic flux for spin $\sigma$ electrons moving around the depicted triangles. It is this nonzero spin dependent flux that causes spin quantum Hall effect for non-interacting electrons.}\label{Fig1}
\end{center}
\end{figure}

For the parameter space defined above and at half filling (undoped case), we can use standard techniques such as canonical transformations or the second order perturbation to obtain an effective Hamiltonian for the spin degree of freedom. Integrating out the hopping to the nearest neighbor term, we obtain the $J1-J2$ Heisenberg model
\begin{equation}
\label{Heisenberg-model}
H_{J_1-J_2}=J_{1}\sum_{\langle ij \rangle} {\rm S}_{i}.{\rm S}_{j}+ J_{2}\sum_{\langle \langle ij \rangle \rangle } {\rm S}_{i}.{\rm S}_{j},
\end{equation}
where $J_1$ and $J_2$ are related to the parameters of the Hubbard model as follows
\begin{equation}
\label{J1-J2}
J_{1}=  4 \frac{t^2}{U} - 16 \frac{t^4}{U^3}   \ld , \ld J_{2}=4\frac{t^4}{U^3}.
\end{equation}

Integrating out the Kane-Mele term leads to the following effective Hamiltonian

\begin{equation}
\label{g2}
H_{g_2}= g_{2}\sum_{\langle \langle ij \rangle \rangle } {\rm S}^{z}_{i}{\rm S}^{z}_{j}-{\rm S}^{x}_{i}{\rm S}^{x}_{j}-{\rm S}^{y}_{i}{\rm S}^{y}_{j},
\end{equation}
where $g_{2}=4\frac{\lambda_{\rm SO}^2}{U}$. This term generates a ferromagnetic interaction for the $x$ and $y$ components and antiferromagnetic interaction for the $z$ component of the spin operator. Considering the antiferromagnetic Heisenberg interaction, $g_2$ adds frustration to the ${\rm S}^{z}{\rm S}^{z}$ interaction and reduces frustration from the in plane ${\rm S}^{x}{\rm S}^{x}+{\rm S}^{y}{\rm S}^{y}$ interaction. Therefore, when $g_2$ term dominates over the $J_2$ Heisenberg interaction, it tends to align spins in the XY plane.

The effective Hamiltonian is the sum of the $H_{J_1-J_2}$ and $H_{g_2}$ as follows
\begin{equation}
\label{Kane-Mele-Hesinberg}
H=J_{1}\sum_{\langle ij \rangle} {\rm S}_{i}.{\rm S}_{j}+ \sum_{\langle \langle ij \rangle \rangle }\left[J_{2,\perp}\para{{\rm S}^{x}_{i}{\rm S}^{x}_{j}+{\rm S}^{y}_{i}{\rm S}^{y}_{j}}+J_{2,\parallel} {\rm S}^{z}_{i}{\rm S}^{z}_{j}\right],
\end{equation}
where
\begin{eqnarray}
\label{Jv-Jz}
&J_{2,\perp}=  J_2-g_2=4 \frac{t^4}{U^3}-4\frac{\lambda_{\rm SO}^2}{U}&\cr
&J_{2,\parallel}=J_2+g_2= 4 \frac{t^4}{U^3}+4\frac{\lambda_{\rm SO}^2}{U}&.
\end{eqnarray}

In order to obtain the phase diagram of the strongly correlated Kane-Mele-Hubbard model, we can invert the above equations to solve $U/t$ and $\lambda_{\rm SO}/t$ in terms of $J_1-J_{2,\perp}-J_{2,\parallel}$ parameters as follows
\begin{eqnarray}
&&\frac{\lambda_{\rm SO}}{t}=\sqrt{\frac{\para{J_{2,\parallel}-J_{2,\perp}}}{2\para{J_1+2J_{2,\parallel}+2J_{2,\perp}}}}\cr
&&\frac{U}{t}=\sqrt{2\frac{J_1+2J_{2,\parallel}+2J_{2,\perp}}{J_{2,\parallel}+J_{2,\perp}}}.
\end{eqnarray}

Since the above Hamiltonian describes only the spin degree of freedom and is a generalization for the Heisenberg model, we name it {\em Kane-Mele-Heisenberg} Hamiltonian. Because of the extra parameter in this model, its phase diagram is expected to be much richer than the phase diagram of the $J_1-J_2$ Heisenberg model which is extensively studied in the literature \cite{Mosadeq_2010_a,Clark_2010_a,Reuther_2011_a,Albuquerque_2011_a,Ying_2010_1,Wang_Fa_2010a}. In this paper, we study the phase diagram of the Kane-Mele-Heisenberg Hamiltonian using several theoretical and numerical approaches. Before presenting formal discussions we would like to comment on the possible phases for the above Hamiltonian based on general arguments. The phase diagram depends on the two ratios, $p_1=J_{\parallel}/{J_1}$ and $p_2=J_{\perp}/J_{\parallel}$. Naively speaking, the next nearest neighbor interaction becomes important when $p_1$ is comparable to one. When $p_1$ is small, we expect a gapless spin liquid phase within the mean field approximation.  On the other hand, for large values of $p_1$, when $p_2 \to 1$ the model is closer to the $J_1-J_2$ Heisenberg model and we can neglect the spin orbit coupling term. For those parameters, we expect a chiral spin liquid phase with gapped spin excitation spectrum which is described by the Haldane model in the mean field approximation. In the opposite limit where $p_2 \to -1$, the Heisenberg interaction is negligible and the spin orbit coupling dominates. In this regime, a topological spin liquid phase with nonzero spin Hall conductance is expected which is described by the Kane-Mele model within the mean field level. In the remainder of this paper we present calculations based on the Schwinger boson/fermion approaches to study the KMH model in more details.

\section{Schwinger bosons approach}

Heisenberg interaction enjoys a global SU(2) spin rotation symmetry. Forgetting about the quantum mechanical nature of the spin operators, a simple classical analysis of the spontaneous symmetry breaking yields the anti-ferromagnetic Neel ordering as the groundstate of the Heisenberg model. However, taking quantum fluctuations may melt the Neel order with solid-like long range order(LRO) down to a spin liquid which lacks LRO and symmetry breaking down to the lowest temperatures. One way to study this quantum transition is the Shwinger boson approach. Spin operators can be written in terms of Schwinger bosons (slave bosons or partons). Bosons exhibit two different behaviors. Their excitation energy is either gapless or gapped. In the gapless excitation case, Shwinger bosons condense at some momentum vector through the Bose-Einstein condensation (BEC). Following the definition of the spin operators in terms of these new Schwinger boson quasiparticles, they acquire a nonvanishing expectation value i.e. magnetic order emerges in the ground-state. The spatial pattern of the magnetic is determined by the momentum at which Shwinger bosons condense.

the second case where Shwinger bosons are gapped and therefore do not undergo BEC transition even at zero temperature, leads to a vanishing expectation value for the spin operators. Subsequently, we obtain a spin liquid phase by definition in that case.

What if we add the spin interaction caused by the spin-orbit interaction? According to the equation [\ref{g2}], the $g_2$ term breaks the global SU(2) spin rotation down to the global U(1) spin rotation around the z-axis. This means within the classical physics, the spin operator has a nonzero in-plane value (XY ordering), however there is still a degree of freedom for choosing the direction of that axis in the XY plane. Again, the quantum mechanics can change the story drastically. Schwinger boson approach can help us to overcome this dilemma in determining the fate of the spin ground-state. In the following, we present more detail on applying the Schwinger fermion approach and extract the phase diagram.

Before staring with the Schwinger boson model, we would like to comment on the idea behind that procedure. To that end, we first employ the {\bf hardcore boson} representation of spin operators, in which an empty site denotes the spin down and an occupied site denotes the spin up states at that site. Therefore, we use the following mappings

\begin{equation}
{\rm S}^{+}_{i} \to d_{i}^\dag \sd,\sd {\rm S}^{-}_{i} \to d_{i} \sd,\sd S^{z}_{i}\to d_{i}^\dag d_{i}- \frac{1}{2},\ld
\end{equation}
where $d_{i}$ is the annihilation operator of hardcore boson at site $i$. It is easy to check that the above definitions recovers the SU(2) group symmetric relations for the spin generator. The dimension of the Hilbert is also two per site as we used hardcore bosons. However, working with hardcore bosons is hard. An acclaimed method is the slave boson where employs two types of bosons, $d_{i,\uparrow}^\dag$ which denotes the creation of the hardcore boson and $b_{i,\downarrow}^\dag$ representing the empty site. Because we either have an occupied or an empty site, we need to implement the $b_{i,\uparrow}^\dag b_{i,\uparrow}+b_{i,\downarrow}^\dag b_{i,\downarrow}=1$ local constraint at every site. Therefore, we use $d_{i}^\dag =b_{i,\uparrow}^\dag b_{i,\downarrow}$ as the slave boson representation of the hardcore boson. It is clear from the definition of the $b_{i}$ along with the local constraint, that we either have $\ket{\uparrow}_i=b_{i}^
\dag \ket{0}=d_{i,\uparrow}^\dag \ket{{\rm vac}}_i$, and $\ket{\downarrow}_i=\ket{0}=d_{i,\downarrow}^\dag \ket{{\rm vac}}_i$, where $\ket{{\rm vac}}$ is an unphysical state which does not host any quasiparticle (neither the empty nor the occupied state by the hardcore boson).

Therefore, in the Schwinger boson approach, we decompose the spin operator in terms of two flavors of bosons ${\bf S}_i=\frac{1}{2}{\bf b}^\dag_{i}{\bf \sigma} {\bf b}_{i}$, where ${\bf b}_{i}=\para{b_{i,\uparrow},b_{i,\downarrow}}^{\rm T}$. We also need to impose the local constraint: $b_{\uparrow,i}^\dag b_{\uparrow,i}+b_{\downarrow,i}^\dag b_{\downarrow,i}=1 $, to recover the physical Hilbert space. Using the Schwinger boson approach, we come to the following relations

\begin{eqnarray}
  &&4{\rm S}_{i}.{\rm S}_{j}=-2\hat{\Delta}_{s,i,j}^\dag \hat{\Delta}_{s,i,j}+1\cr
  &&=1+2\hat{\chi}_{s,i,j}^\dag \hat{\chi}_{s,i,j}=\hat{\chi}_{s,i,j}^\dag \hat{\chi}_{s,i,j}-\hat{\Delta}_{s,i,j}^\dag \hat{\Delta}_{s,i,j}\\
  &&4{\rm S}_{i}.{\rm \tilde{S}}_{j}=-2\hat{\Delta}_{t,i,j}^\dag \hat{\Delta}_{t,i,j}+1\cr
  &&=1+2\hat{\chi}_{t,i,j}^\dag \hat{\chi}_{t,i,j}=\hat{\chi}_{t,i,j}^\dag \hat{\chi}_{t,i,j}-\Delta_{t,yi,j}^\dag \hat{\Delta}_{t,i,j},
\end{eqnarray}
in which $\hat{\chi}_{s,i,j}=~b_{i,\uparrow}^\dag b_{j,\uparrow}+b_{i,\downarrow}^\dag b_{j,\downarrow}$, $\hat{\chi}_{t,i,j}=~b_{i,\uparrow}^\dag b_{j,\uparrow}-b_{i,\downarrow}^\dag b_{j,\downarrow}$, $\hat{\Delta}_{s,i,j}=~b_{i,\uparrow}b_{j,\downarrow}-b_{i,\downarrow}b_{j,\uparrow}$, and $\hat{\Delta}_{t,i,j}=~b_{i,\uparrow}b_{j,\downarrow}+b_{i,\downarrow}b_{j,\uparrow}$. Therefore the second term in the Kane-Mele-Heisenberg Hamiltonian which is $g_{2}{\rm S}_{i}.{\rm \tilde{S}}_{j}+J_{2}{\rm S}_{i}.{\rm S}_{j}$ decouples as
\begin{eqnarray}
  &&-\frac{1}{2}J_{2}\hat{\Delta}_{s,i,j}^\dag \hat{\Delta}_{s,i,j}-\frac{1}{2}g_{2}\hat{\Delta}_{t,i,j}^\dag \hat{\Delta}_{t,i,j}.\sd
\end{eqnarray}

It is worthwhile mentioning that $\hat{\Delta}_{s,j,i}=-\hat{\Delta}_{s,i,j}$, while $\hat{\Delta}_{t,j,i}=\hat{\Delta}_{t,i,j}$. Fa Wang in Ref. \cite{Wang_Fa_2010a} has studied this 
model at $g_2=0$ using the Schwinger bosons. In the following we closely follow him and extend his study to the nonzero $g_2$ values. To study the above model formally, we appeal to the Hubbard-Stratonovic transformation followed by the saddle point approximation. Let us use the following definitions
\begin{eqnarray}
  & \mu=&\braket{\lambda_i}\cr
  &\Delta_{s,i,j}=&\braket{\hat{\Delta}_{s,i,j}}\cr
  &\delta_{t,i,j}=&\braket{\hat{\Delta}_{t,i,j}}.
\end{eqnarray}

We also assume $\chi_{i,j}=\braket{\tilde{\chi}}_{i,j}=0$. To obtain the mean field Hamiltonian we assume the zero flux state pattern for $\Delta_{1,s}$ and $\Delta_{2,s}$ parameters which is introduced in Ref. \cite{Wang_Fa_2010a}. We also assume a uniform s-wave $\Delta_{2,t}$ in the mean field state. Accordingly, the mean field Hamiltonian can be rewritten as follows
\begin{widetext}
\begin{eqnarray}
H_{MF}=&&\sum_{k} \Psi_{k}^\dag \left(
        \begin{array}{cccc}
          \mu & 0 & \Delta_{2,k} & \eta_{k} \\
          0 & \mu & -\eta_{k}^{*} & \Delta_{2,k} \\
          \Delta_{2,k}^{*} & -\eta_{k} & \mu & 0 \\
          \eta_{k}^{*} & \Delta_{2,k}^{*} & 0 & \mu \\
        \end{array}
      \right) \Psi_{k}+N_{s}\para{\frac{3}{2}J_{1}\Delta_{1,s}^2+3J_{2}\Delta_{2,s}^2+3g_{2}\Delta_{2,t}^2+\mu},\end{eqnarray}
\end{widetext}
where $N_s$ is the number of sites, $\Psi_{k}=\para{b_{k,A},b_{k,B},b_{-k,A}^\dag,b_{-k,B}^\dag}^{{\rm T}}$, and
\begin{eqnarray}
&&\eta_{k}=\frac{J_{1}}{2}\Delta_{1,s}\para{\exp\para{-ik_y}+2\exp\para{ik_y/2}\cos\para{\frac{\sqrt{3}}{2}k_x}},\cr
&&\Delta_{2,k}=\xi_{t,k}+i\xi_{s,k},\cr
&&\xi_{s,k}=2J_{2}\Delta_{s,t}\sin\para{\sqrt{3}k_x/2}\para{\cos\para{3k_y/2}-\cos\para{\sqrt{3}k_x/2}},\cr
&&\xi_{t,k}=g_{2}\Delta_{2,t}\para{2\cos\para{\sqrt{3}k_x/2}\cos\para{3k_y/2}+\cos\para{\sqrt{3}k_x}}.\ld
\end{eqnarray}

The above Hamiltonian can be diagonalized using the Bogoliubov transformations which reduces to finding the eigenvalues of the $M_{k}{\Lambda}$ matrix where $\Lambda={\rm diag}\para{1,1,-1,-1}$. Accordingly, the energy dispersion has two branches as follows

\begin{equation}
  E_{k}^{\pm}=\sqrt{\mu^2\pm 2\abs{\eta_{k}}\xi_{s,k}-\abs{\eta_k}^2-\xi_{s,k}^2-\xi_{t,k}^2}.
\end{equation}

Minimizing the total energy
\begin{equation}
  E_{\rm tot}=N_{s}\para{\frac{3}{2}J_{1}\Delta_{1,s}^2+3J_{2}\Delta_{2,s}^2+3g_{2}\Delta_{2,t}^2+\mu}+\sum_{k} E^{\pm}_k ,
\end{equation}
with respect to $\mu$, $\Delta_{1,s}$, $\Delta_{2,s}$, and $\Delta_{2,t}$, we obtain the phase diagram of the KMH model. Among the self-consistency equations emerging from the minimization with respect to $\mu$ we have the following constraint
\begin{eqnarray}
  \sum_{k}\para{\frac{\abs{\mu}}{E^{+}_{k}}+\frac{\abs{\mu}}{E^{-}_{k}}-2}=0,
\end{eqnarray}
that implements the local constraint on the Hilbert space in average. To achieve the phase diagram we need to determine whether or not the energy excitation of Schwinger bosons i.e. $E^{\pm}_k$ is gapped. If gapped we obtain the spin liquid phase, while the gapless case corresponds to the magnetic ordering. If bosons condense in site $i$ condense such that $\braket{b_{i,\uparrow}}=z_{1,i}$, and $b_{i,\downarrow}=z_{2,i}$, we have

\begin{eqnarray}
&&  \braket{S_x}\para{i}=\Re \para{z_{1,i}^{*}z_{2,i}}, \cr
&&  \braket{S_y}\para{i}=\Im \para{z_{1,i}^{*}z_{2,i}},\cr
&&  \braket{S_z}\para{i}=\frac{1}{2}\para{\abs{z_{1}}^2-\abs{z_{2}}^2}.
\end{eqnarray}

We also need to find at which momentum Schwinger bosons condense. For example if we obtain $\Delta_{2,t}=0$ and Schwinger bosons condense at $K=\para{0,0}$ i.e. $E_{0,0}^{\pm}=0$, we have $\mu=-\eta_{k=\para{0,0}}=-\frac{3J_1}{2}$. The eigenvectors corresponding to the zero mode are given by $\para{1,0,0,-1}$ and $\para{0,1,1,0}$. Assuming the weight of Shwinger bosons that condense at $E^{+}_{0,0}$ is $z_{1}$ and those at $E^{-}_{0,0}$ is $z_{2}$ we have

\begin{eqnarray}
\braket{\Psi_{0,0}}^{\rm T}=\para{\braket{b_{A,\uparrow}},\braket{b_{B,\uparrow}},\braket{b_{A,\downarrow}}^*,\braket{b_{B,\downarrow}}^*}=\para{z_1,z_2,z_2,-z_1},~\sd
\end{eqnarray}
which means
\begin{eqnarray}
&&\braket{S^{z}_A}= \frac{1}{2}\para{\abs{\braket{b_{A,\uparrow}}}^2-\abs{\braket{b_{A,\downarrow}}}^2}=\frac{1}{2}\para{\abs{z_1}^2-\abs{z_2}^2},\cr
&&\braket{S^{z}_B}= \frac{1}{2}\para{\abs{\braket{b_{B,\uparrow}}}^2-\abs{\braket{b_{A,\downarrow}}}^2}=\frac{1}{2}\para{\abs{z_2}^2-\abs{z_1}^2}=-\braket{S^{z}_A}.\cr
&&\braket{S^{+}_A}= \braket{b_{A,\uparrow}}^*\braket{b_{A,\downarrow}}= z_1^*z_2,\cr
&&\braket{S^{+}_B}=\braket{b_{B,\uparrow}}^*\braket{b_{B,\downarrow}}=- z_2z_1^*=-\braket{S^{+}_A}.
\end{eqnarray}

Therefore the aforementioned condition leads to the Neel spin ordering. In the most general case, when bosons condense at ${\bf k}$, the spatial profile of magnetization is a wave-packet with wave-vector ${\bf k}$. That means $\braket{{\bf \rm S}}_{j,\tau}=\exp\para{ik.R_{ji}}\braket{{\bf \rm S}}_{i,\tau}$, where $R_{ji}=R_{j}-R_{i}$ and $\tau=\left\{A,B\right\}$. In case ${\bf k}$ is commensurate, i.e. ${\bf k}=\frac{p}{q}{\bf G}_{1,2}$, where $p,q$ are two integers with no common devisor and ${\bf G}_{1,2}$ are basis vectors of the reciprocal lattice, the unit cell encloses $2\abs{q}$ sites. An interesting case is ${\bf k}={\bf K}$ and ${\bf k}={\bf K'}$, where $K$ and $K'$ are those momenta at which $\eta_{k}=0$. Since $3{\bf K}=3{\bf K'}\equiv {\bf G}_{1,2}$, the unit cell triples. Bosons condense at these momenta when $J_2\to \infty$  where we obtain two decoupled triangular lattices, and it is well known that magnetic ordering  triples the unit cell and spins form 120 degrees with neighboring spins. It should be noted that the minimizing the total energy of bosons is somewhat tricky and there is small numerical error. Therefore, we have presented a schematic phase diagram in Fig. [2] based on our numerical results.

\begin{figure}[tbp]
\begin{center}
\includegraphics[width=\linewidth]{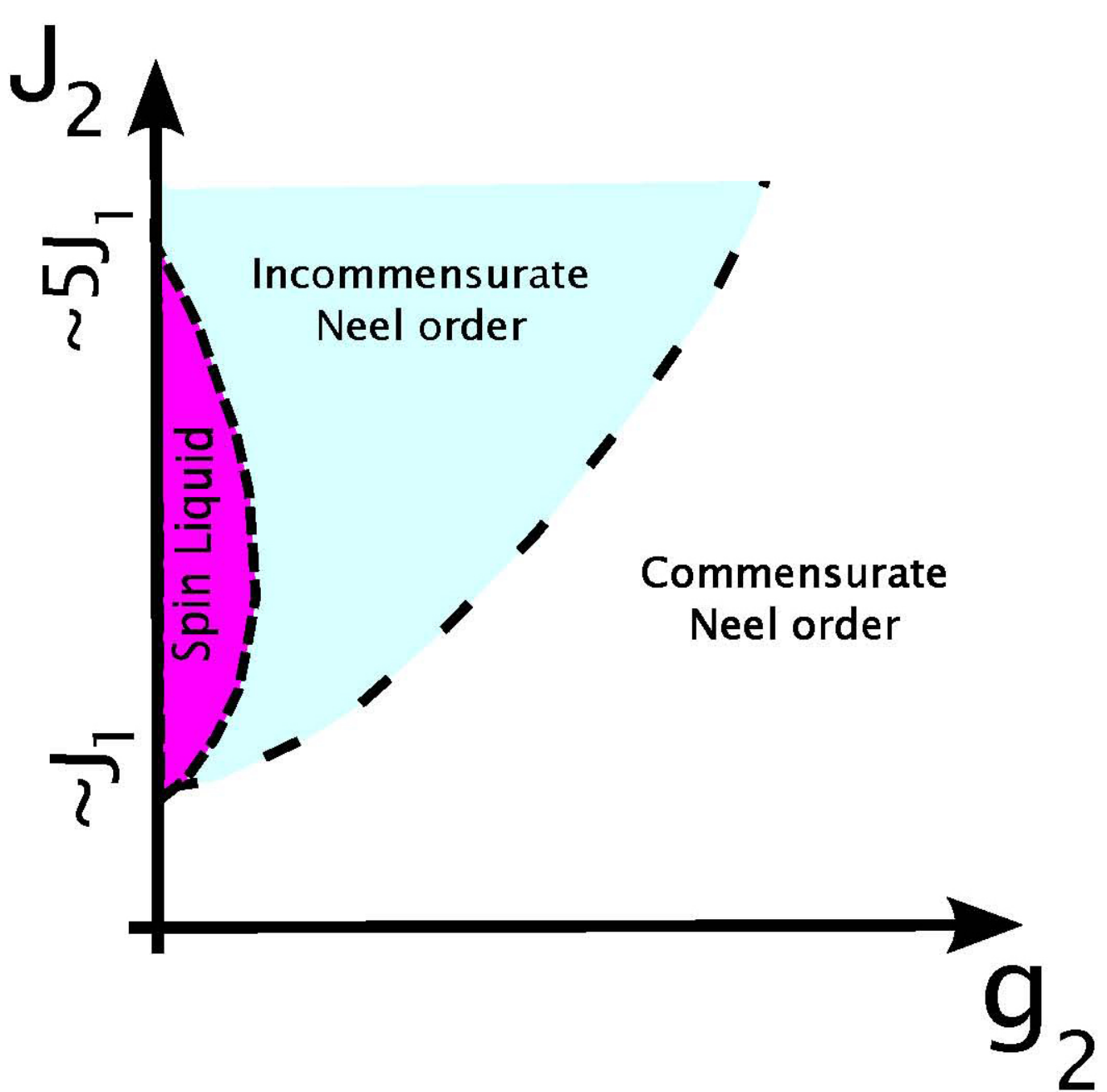}
\caption{(Color online) Schematic mean field phase diagram of the KMH using the Schwinger model based on the numerical minimization of the total energy.}\label{Fig2}
\end{center}
\end{figure}

\section{Schwinger fermion approach}

As we discussed in the previous section, Schwinger boson approach is a useful tool to identify the spin liquid phase in the phase diagram. The mean field Schwinger boson method sparks the existence of the gapped spin liquid phase for the intermediate values of $J_2/J_1$ and for small values of $g_2/J_2$. The rest of the phase diagram is prone to exhibit magnetism of either commensurate or incommensurate Neel ordering forms. In this section, we are going to study the spin liquid phase more carefully. To that end, we employ the Schwinger fermion approach to represent the spin operators. The decomposition procedure for the spin operators is the same as that of the Schwinger bosons expect that we replace bosonic $b_{i,\sigma}$ operators with fermionic ones $f_{i,\sigma}$. In the following, we consider the competition between spin liquids only as the starting point, i.e. we assume there is no long range magnetic ordering in the ground-state. However, after discussing the gauge theory of the KMH model, we argue that only the gapped spin liquid obtained in the previous section is stable against gauge fluctuations such as instanton effect for certain and other proposed spin liquid are likely unstable toward spontaneously broken symmetry phases. A key result is that the gapped spin liquid phase is a chiral spin liquid, i.e. the spin excitation in the bulk is gapped, while it has topologically protected gapless edge modes. This result is consist with several previous studies \cite{He_2011_a,Fiete_2011_b,Lu_2010_a}.

The spin 1/2 operator can be represented in terms of Schwinger fermions with two flavors subject to a constraint (such that the dimension of the local Hilbert space be two as that of the spin 1/2) as follows
\begin{eqnarray}
\label{S-F}
&&S^{+}_{i}=f_{i,\uparrow}^\dag f_{i,\downarrow}\sd, \sd S^{-}_{i}=f_{i,\downarrow}^\dag f_{i,\uparrow}\sd,\sd S^{z}_{i}=\frac{n_{i,\uparrow}-n_{i,\downarrow}}{2}\cr
&& n_{i,\uparrow}+n_{i,\downarrow}=f_{i,\uparrow}^\dag f_{i,\uparrow}+f_{i,\downarrow}^\dag f_{i,\downarrow}=1 .
\end{eqnarray}

Due to the constraint $S^{z}_{i}$ operator can also be written as $S^{z}_{i}=n_{i,\uparrow}-1/2=1/2-n_{i,\downarrow}$. Using the Schwinger fermion redefinition of spin operator we can rewrite the Kane-Mele-Heisenberg Hamiltonian in the following way

\begin{eqnarray}
\label{S-F-KMH}
H=~&&-J_1/2~ \sum_{\langle i,j \rangle} ~~\hat{\chi}^\dag \para{i,j} \hat{\chi}\para{i,j} \cr
&&-J_{\perp}/2  \sum_{\langle \langle i,j \rangle \rangle}\para{\hat{\chi}_{\uparrow}^{\dag}\para{i,j}  \hat{\chi}_{\downarrow}\para{i,j} +H.c.} \cr
&&-J_{\parallel}/2 \sum_{\langle\langle i,j\rangle \rangle} \para{\hat{\chi}_{\uparrow}^{\dag}\para{i,j} \hat{\chi}_{\uparrow}\para{i,j}+\hat{\chi}_{\downarrow}^{\dag}\para{i,j} \hat{\chi}_{\downarrow}\para{i,j}},~\sd
\end{eqnarray}
in which we have defined $\hat{\chi}_{\sigma}\para{i,j}=f_{i,\sigma}^\dag f_{j,\sigma}$ and $\hat{\chi}\para{i,j}=\hat{\chi}_{\uparrow}\para{i,j}+\hat{\chi}_{\downarrow}\para{i,j}$. The constraint can be implemented using the Lagrange multiplier method, though at half filling, it can be set to zero at the level of mean field approximation. By implementing a Hubbard-Stratonovic transformation, the above Hamiltonian can be approximated by the following effective Hamiltonian

\begin{eqnarray}
\label{S-F-KMH}
&&H_{MF}=-\frac{1}{2} ~\sum_{\langle i,j \rangle,\sigma} ~J_1\chi_{1} \para{i,j} f_{i,\sigma}^\dag f_{j,\sigma} \cr
&&-\frac{1}{2} \sum_{\langle \langle i,j \rangle \rangle,\sigma}\para{J_{\perp}\chi_{2,-\sigma}\para{i,j}^{*}+J_{\parallel}\chi_{2,-\sigma}\para{i,j}^{*}}f_{i,\sigma}^\dag f_{j,\sigma}~~~
\end{eqnarray}
where $\chi_{\sigma}\para{i,j}=\braket{\hat{\chi}_{\sigma}\para{i,j}}$. To obtain the phase diagram, we assume $\chi_{1}\para{i,j}=\chi_{1}$ and $\chi_{2,\sigma}\para{i,j}=\chi_{2} \exp\para{i\phi_{\sigma}\nu_{i,j}}$ in which we have followed the convention for $\nu_{i,j}$ used in the Kane-Mele-Hubbard model (see equation [\ref{KMH-model}] for example). Using these assumptions the Hamiltonian becomes quadratic and is described by the following Matrix Hamiltonian for each spin degree of freedom
\begin{widetext}
\begin{eqnarray}
\label{S-F-Matrix-1}
&&H_{MF}=\sum_{k,\sigma} \para{\begin{array}{cc}
                           f_{k,A,\sigma}^\dag & f_{k,B,\sigma}^\dag
                         \end{array}} \left(
                                                                                                               \begin{array}{cc}
                                                                                                                  \tilde{\xi}_{k,\sigma}+\bar{\xi}_{k,\sigma}& \eta_{k} \\
                                                                                                                 \eta_{k}^{*} & \tilde{\xi}_{k,\sigma}-\bar{\xi}_{k,\sigma} \\
                                                                                                               \end{array}
                                                                                                             \right)
                         \para{\begin{array}{c}
                            f_{k,A,\sigma} \\
                            f_{k,B,\sigma}
                          \end{array}}
\end{eqnarray}
\end{widetext}
where
\begin{eqnarray}
\label{S-F-Matrix-2}
&&\eta_{k}=-J_{1}\chi_1\para{\exp\para{-ik_y}+2\cos\para{\sqrt{3}k_x/2}\exp\para{ik_y/2}}\sd~\\
&&\bar{\xi}_{k,\sigma}=-\chi_2\para{J_{2,\parallel}\sin{\phi_{\sigma}}+J_{2,\perp}\sin{\phi_{-\sigma}}}\bar{\zeta}_{k}\cr
&&\bar{\zeta}_{k}=2\sin\para{\sqrt{3}k_x/2}\para{\cos\para{3k_y/2}-\cos\para{\sqrt{3}k_x/2}}\sd\\
&&\tilde{\xi}_{k,\sigma}=-\chi_2\para{J_{2,\parallel}\cos{\phi_{\sigma}}+J_{2,\perp}\cos{\phi_{-\sigma}}}\tilde{\zeta}_{k}\cr
&&\tilde{\zeta}_{k}=2\cos\para{\sqrt{3}k_x/2}\cos\para{3k_y/2}+\cos\para{\sqrt{3}k_x},\sd
\end{eqnarray}
and the energy spectrum is given by

\begin{equation}
\label{S-F-Energy}
E_{k,\sigma}=\tilde{\xi}_{k,\sigma}\pm \sqrt{\abs{\eta_{k}}^2+\bar{\xi}_{k,\sigma}^2}.
\end{equation}

At half filling all energy levels in the lower bands are occupied by Schwinger fermions. There are four parameters in the total energy, $\chi_1$, $\chi_2$, $\phi_{\uparrow}$ and $\phi_{\downarrow}$. To obtain their optimum values at half filling, the selfconsistency equations for $\chi_{\sigma}\para{i,j}=\braket{f_{i,\sigma}^\dag f_{j,\sigma}}$ should be solved. Alternatively, we can minimize the total energy with respect to those parameters and we obtain the same values. We have solved the selfconsistency equations numerically and we obtain three different phases, 1- {\em gapless spin liquid}, 2- {\em chiral gapped spin liquid}, with nonzero Hall conductance and protected gapless edge states, and 3- {\em topological gapped spin liquid}, with nonzero spin Hall conductance and protected gapless edge states (see Fig. [3] for the phase diagram). In the following we present more details on the nature of these phases.

\begin{figure}[tbp]
\begin{center}
\includegraphics[width=\linewidth]{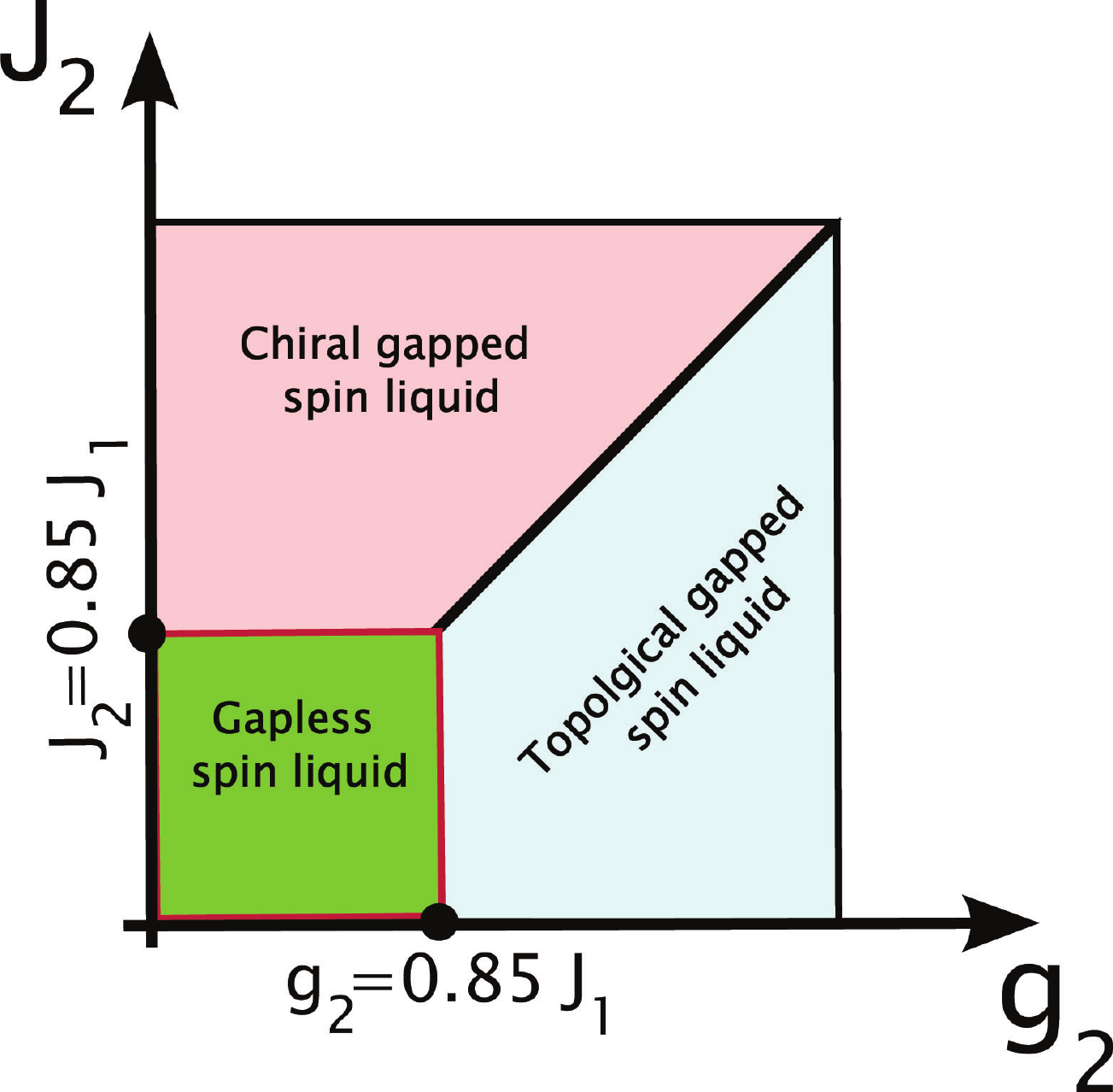}
\caption{(Color online) Mean field phase diagram of the KMH within the Schwinger fermion model.}\label{Fig3}
\end{center}
\end{figure}

{\bf 1- Gapless spin liquid phase.---} Numerical minimization of the total energy shows that when $J_{2,\parallel}+\abs{J_{2,\perp}}\leq 1.7 J_{1}$ (or equivalently $J_2<.85J_1$ and $g_2<.85J_1$), the optimum value for $\chi_{2}$, $\phi_{\uparrow}$, and $\phi_{\downarrow}$ are all equal to zero. In this phase, the energy dispersion of Schwinger fermions is identical to the energy dispersion of electrons in noninteracting graphene sheets with $t_{\rm eff}=J_{1}\chi_{1}$. Accordingly, the spin excitation (i.e. the excitation energy for Schwinger fermions) is gapless. On the other hand, it is easy to check that at the mean field level, there is no long range spin or charge ordering. Therefore this gapless phase does not break any lattice symmetry and by definition is a gapless spin liquid phase. This result is however a consequence of the mean field approximation and needs to be more carefully studied under fluctuations around the mean field ground-state which is a result of the strong correlations among Schwinger fermions. For example fluctuations of mean field parameters i.e. $\chi_{1,2}$ and $\phi_{\uparrow,\downarrow}$ may destroy the properties of the mean field state. It has been discussed in the literature that the most important fluctuations which are should be included in any serious study are compact gauge fluctuations. In the next section we discuss that this state may undergo phase transition to the anti-ferromagnetic or VBS state \cite{Vaezi_2011_a,Fu_2010_a}.

{\bf 2- Chiral gapped spin liquid phase.---} For $J_{2,\parallel}+\abs{J_{2,\perp}}\geq 1.7 J_{1}$ and $J_{2,\perp}>0$ (or equivalently $J_2>.85J_1$ and $J_2>g_2$), the minimum of the ground-state energy manifold yields nonzero values for both $\chi_1$ and $\chi_{2}$ and we obtain $\phi_{\uparrow}=\phi_{\downarrow}=\pm \frac{\pi}{2}$ and the effective Hamiltonian for the Schwinger fermions within the mean field approximation is given by
\begin{eqnarray}
\label{Haldane-1}
H=&&-J_{1}\chi_{1} \sum_{\langle ij \rangle,\sigma} f^{\dag}_{i,\sigma}f_{j,\sigma}\cr
&&+ i\para{J_{2,\parallel}+J_{2,\perp}}\chi_{2}\sum_{\langle\langle i,j \rangle\rangle,\sigma}\nu_{ij}f^{\dag}_{i,\sigma}f_{j,\sigma},
\end{eqnarray}
which is identical to the Haldane model for the chiral state on the honeycomb lattice. Consequently, the magnetic flux penetrating triangulares consisting of three neighboring same-sublattice sites equals $\bar{\Phi}_{\sigma}=\pm3\phi_{\uparrow}\neq 0$ for both $f_{\uparrow}$ and $f_{\downarrow}$ Schwinger fermions. The above model in the continuum model is described by two gapped Dirac cones around $\vec{K}=$ and $\vec{K'}=-\vec{K}$ and the Chern of each band equals $C^{\sigma}={\rm sgn}\para{\phi_{\uparrow}}$ regardless of the flavor(spin) of the Schwinger fermions. Therefore the Hall conductance of the system is $\sigma_{xy}^{\uparrow,\downarrow}=C^{\uparrow,\downarrow}\frac{{\rm e}^2}{2{\rm hc}}={\rm sgn}\para{\phi_{\uparrow}}\frac{{\rm e}^2}{{\rm hc}}$. This means that the density of Schwinger fermions changes by $\Delta n_{\sigma}={\rm sgn}\para{\phi_{\uparrow}}\frac{{\rm e}^2}{{\rm hc}} \Phi$ after inserting $\Phi$ magnetic flux (which couples to Schwinger fermions). Since $\abs{C^{\sigma}}=1$, there is one protected chiral edge mode for each spin degree of freedom. These chiral edge modes are robust against disorder and their chirality is given by the sign of $C^{\sigma}$. Since $\sum_{\sigma} \bar{\Phi}_{\sigma}\neq \{0,\pi\}$, the system breaks the time reversal symmetry. This can also be seen in the spin chirality value. Spin chirality is defined as ${\rm E}_{i,j,k}=\braket{{\rm S_{i}. (S_{j}\times S_{k})}}=2i\braket{\hat{\chi}_{i,j}\hat{\chi}_{j,k}\hat{\chi}_{k,i}-\hat{\chi}_{i,k}\hat{\chi}_{k,j}\hat{\chi}_{j,i}}$, where $\chi_{i,j}=\sum_{\sigma}f_{i,\sigma}^\dag f_{j,\sigma}$. The spin chirality operator is odd under the time reversal symmetry $T$, and parity $P$. Condensation of ${\rm E}_{i,j,k}$ implies the breaking of the time reversal and parity symmetries in the ground state. In this phase, it can be shown that ${\rm E}_{i,j,k} \propto \abs{\bar{\Phi}}$ when $i,j,k$ belong to the same sublattice and are nearest neighbor of each other (belong to a triangle). However, within the meanfield level, the ground state respects the spin rotation, $C_{6}$(120 degrees rotation symmetry), and the translational symmetries. Therefore we name it chiral gapped spin liquid phase.

{\bf 3- Topological gapped spin liquid phase.---} For $J_{2,\parallel}+\abs{J_{2,\perp}}\geq .85 J_{1}$ and $J_{2,\perp}<0$ (or equivalently $g_2>.85J_1$ and $g_2>J_2$), we obtain both $\chi_1$ and $\chi_{2}$ nonzero and $\phi_{\sigma}=\sigma \phi=\pm \sigma \frac{\pi}{2}$. The effective Hamiltonian for the Schwinger fermions within the mean field approximation is given by
\begin{eqnarray}
\label{K-M-1}
H=&&-J_{1}\chi_{1} \sum_{\langle ij \rangle,\sigma} f^{\dag}_{i,\sigma}f_{j,\sigma}\cr
&&+ i\para{J_{2,\parallel}+J_{2,\perp}}\chi_{2}\sum_{\langle\langle i,j \rangle\rangle,\sigma}\sigma \nu_{ij}f^{\dag}_{i,\sigma}f_{j,\sigma},
\end{eqnarray}
which is identical to the Kane-Mele model for the topological insulator phase on the honeycomb lattice. Consequently, the magnetic flux penetrating triangulares consisting of three neighboring same-sublattice sites equals $\bar{\Phi}_{\uparrow}=\pm 3\phi_{\uparrow}$ for $f_{\uparrow}$ and $\bar{\Phi}_{\downarrow}=-\bar{\Phi}_{\uparrow}$ for $f_{\downarrow}$ Schwinger fermions. The above model in the continuum model reduces to two gapped Dirac cones around $\vec{K}=\frac{4\pi}{3\sqrt{3}}\para{1,0}$ and $\vec{K'}=-\vec{K}$ and the Chern of each band equals $C^{\sigma}=\sigma C=\sigma {\rm sgn}\para{\phi}$ depending on the spin of the Schwinger fermions. Therefore the Hall conductance of the system is $\sigma_{xy}^{\uparrow,\downarrow}=\sigma C\frac{{\rm e}^2}{2{\rm hc}}=\sigma {\rm sgn}\para{\phi}\frac{{\rm e}^2}{{\rm hc}}$. This means that the density of Schwinger fermions changes by $\Delta n_{\sigma}=\sigma {\rm sgn}\para{\phi}\frac{{\rm e}^2}{{\rm hc}} \Phi$ after inserting $\Phi$ magnetic flux (which couples to Schwinger fermions). Therefore the total number density of fermions i.e. $\Delta n_{\uparrow} + \Delta n_{\downarrow} =0$, while the spin density i.e. $1/2 \para{\Delta n_{\uparrow} - \Delta n_{\downarrow}}={\rm sgn}\para{\phi}\frac{{\rm e}^2}{{\rm hc}} \Phi$ is nonzero. Accordingly, the spin Hall conductance of the system is nonzero while fermions do not transfer across the edge of the system. Since $C_{\sigma}=\sigma C$ and $\abs{C}=1$, there is one protected chiral edge mode for each spin degree of freedom with opposite chiralities. The system does not break the time reversal(TR) symmetry and any TR preserving perturbation cannot destroy them due to Kramer's degeneracy. Therefore chiral edge modes are robust against non-magnetic disorder and their chirality is given by their spin and $C$.

\section{Gauge theory of the Kane-Mele-Heisenberg model}

The Kane-Mele-Heisenberg model at half filling is described in terms of spin operators only. As we showed in section IV, these spin operators can be represented in terms of Schwinger fermions. It is straightforward to check that $S_{i}^{x},$ $S_{i}^{y}$, and $S_{i}^{z},$ are all invariant under the following local SU(2) gauge transformations \cite{Affleck_1988_a}

\begin{eqnarray}
&&  f_{i,\uparrow} ~\to ~\alpha_{i} f_{i,\uparrow} +~ \beta_{i} f_{i,\downarrow}^\dag \cr
&&  f_{i,\downarrow}^\dag \to -\beta_{i}^{*} f_{i,\uparrow} + \alpha_{i}^{*} f_{i,\downarrow}^\dag.
\end{eqnarray}

One way to see this is to consider the following matrix

\begin{equation}
\label{SU(2)-1}
\psi_{i}=\left(
  \begin{array}{cc}
    f_{i,\uparrow} & f_{i,\downarrow} \\
    f_{i,\downarrow}^\dag & -f_{i,\uparrow}^\dag \\
  \end{array}
\right).
\end{equation}

Spin operators can be described in terms of $\psi_{i}$ in the following way
\begin{equation}
{\rm {\bf S}}_{i}=\frac{1}{4}\mathrm{Tr}\para{\psi_{i}^\dag \psi_{i} {\bf \sigma}^{\rm T} },
\end{equation}
where ${\bf \sigma}^{\rm T}$ is the transpose of Pauli matrices, ${\bf \sigma}$. From the above definitions, it is obvious that spin operators under $\psi_{i}\to h_{i} \psi_{i}$ where $h_{i}$ is a SU(2) unitary transformation. This transformation is equivalent to the transformation introduced in equation [\ref{SU(2)-1}] provided $h_{11}=\alpha_{i}$ and $U_{12}=\beta_{i}$. Beside the Hamiltonian, the action should also be invariant under SU(2) gauge transformations. To that end, we only need to show that the local constraints on the Hilbert space are also gauge invariant. SU(2) group has three generators, so there should be three constraints to be implemented through three temporal gauge fields that can serve as Lagrange multipliers. At half filling, at any site the total number of Schwinger fermions should by exactly one in order to retain the physical Hilbert space with two states per site. Therefore one constraint is $f_{i,\uparrow}^\dag f_{i,\uparrow}+f_{i,\downarrow}^\dag f_{i,\downarrow}=1$. Two other constraints at half filling can be chosen as $f_{i,\uparrow}^\dag f_{i,\downarrow}^\dag=f_{i,\downarrow} f_{i,\uparrow}=0$ which are direct results of the first constraint. These constraints can be written as  $ \psi_{i}\psi_{i}^\dag=1$ or equivalently

\begin{equation}
\mathrm{Tr}\para{\psi_{i}^\dag \sigma^{\mu}\psi_{i}}=0.
\end{equation}

The above constraint is manifestly gauge dependent. These constraints can be implemented using ${\bf A_{0}}$ gauge fields. Therefore the Lagrangian is

\begin{equation}
\mathcal{L}=\frac{1}{2}\mathrm{Tr}\para{\psi_{i}^\dag \para{i\frac{d}{dt}+\sigma_{\mu} A_{0}^{\mu}\para{i}}\psi_{i}}-H,
\end{equation}

It is straightforward to check that the Lagrangian is invariant under the following simultaneous transformations: $\psi_{i}\to h_{i} \psi_{i}$, and ${\bf A_{0}}\para{i}{\bf .\sigma} \to  h_{i}\left[{\bf A_{0}}\para{i}{\bf .\sigma}+i\frac{d}{dt}\right]h_{i}^\dag$. This completes our claim of the SU(2) gauge invariance of the Kane-Mele-Heisenberg Hamiltonian.

In the above paragraph, we showed that the original model has local SU(2) gauge symmetry. However, the meanfield solution of the system does not necessarily respect this property and can  break the gauge symmetry down to U(1) or $Z_2$ symmetries by the Anderson-Higgs mechanism. To determine the gauge theory of the meanfield state, we need to identify the invariant gauge group~(IGG) of the meanfield solutions. To do so, we investigate the transformation properties of the $\braket{\psi_{i}\psi_{j}^\dag}$ matrices for every $i$ and $j$ sites, under the global SU(2) gauge transformations. When gauge particles are massless, these operators are invariant under any global gauge symmetries and the gauge fluctuations around the meanfield state is described by the original compact SU(2) gauge theory. When only a U(1) subgroup of the SU(2) gauge transformations, leaves all operators intact, the gauge theory is given by a compact U(1). $Z_2$ spin liquid phase is also given by a meanfield ansatz that is invariant under a $Z_2$ subgroup of the SU(2) gauge transformations only. It is easy to show that the IGG of the gapless spin liquid is $SU(2)$ and the other two spin liquid phases have U(1) IGG and therefore we need to consider a spin liquid phase coupled to a compact SU(2) or U(1) gauge theories. Compact gauge theories are in principle hard to study as they include nonperturbative phenomena such as instanton effects~(monopole configurations). Instanton~(anti-instanton) change the value of the gauge potential by $2\pi$~($-2\pi$) and can potentially cause phase transition to symmetry breaking phases such as spin ordered phase, VBS or dimerized state and etc \cite{Vaezi_2011_a,Fu_2010_a}.

\subsection{Gauge theory of the gapless spin liquid phase}
In this phase, the gapless Dirac fermions at $K$ and $K'$ points are coupled to a compact SU(2) gauge fluctuations (${\rm QCD}_3$). The spectrum of the gapless spin liquid phase is similar to the spectrum of the staggered flux phase. The instanton effect on the staggered flux has been extensively studied. Large N expansions indicates that $QCD_3$ can be in the deconfined phase and therefore it might not undergo confinement transition \cite{Hermele_2004_a}. Therefore the gapless spin liquid phase on the honeycomb lattice can be stable against gauge fluctuations and might remain physical even after the instanton proliferation. It should be noted that it is not clear enough whether the large N studies are applicable to the physical SU(2) case. If we believe the other scenario where instantons destabilize the mean field state \cite{Seradjeh_2003_a}, instanton condensation spontaneously breaks lattice symmetries \cite{Vaezi_2011_a,Fu_2010_a}. For instance, we may obtain the VBS state or the Neel order as the ground-state of the KMH in this regime.

\subsection{Gauge theory of the chiral gapped spin liquid phase}

In the chiral gapped spin liquid phase, the meanfield state is described by the gapped Dirac fermions coupled to a compact U(1) gauge field. To obtain an effective action for the gauge field, we can integrate out fermions. Because of the gap in the spectrum of Schwinger fermions, we can easily obtain the Chern number for their energy band. Doing so, we obtain the total Chern number equal to $C=\pm2$ depending on the sign of $\phi_{\uparrow}$. Therefore other than instantons, the U(1) gauge field contains a Chern-Simons terms in addition to the Maxwell action as follows

\begin{eqnarray}
  S_{\rm G.F.}=\int d^2x dt ~\frac{C}{4\pi}\epsilon^{\mu\nu\lambda}a_{\mu}\partial_{\nu}a_{\lambda}-\frac{1}{2{\rm e}^2}\para{\epsilon^{\mu\nu\rho}\partial_{\nu}a_{\lambda}}^2\sd.
\end{eqnarray}

It is well-known that the Chern-Simon term gaps out the gauge particles and therefore, we can neglect the gauge fluctuations as well as the instanton effect in the low energy description. Accordingly, the chiral gapped spin liquid is stable against gauge fluctuations and remains physical.

We would like to mention that the ground state degeneracy of the chiral spin liquid phase is four on the torus. To see why this happens to be true, we should notice that the chiral spin liquid breaks both time reversal and parity symmetries, so the degeneracy equals $2\times2=4$. Another way to check this is to note that fixing the sign of the spin chirality, e.g. $E_{i,j,k}>0$, the ground state degeneracy on the 2D torus is given by the value of the Chern number. Since $\abs{C}=2$, the ground state degeneracy for each $E_{i,j,k}>0$ sector is two. Similarly the ground state degeneracy for the other sector is two as well. Therefore the total degeneracy is four on the torus. This topological invariant serves as good probe to identify the spin liquid phase in numerical techniques such as quantum Monte Carlo (QMC) method.
\begin{figure}[tbp]
\begin{center}
\includegraphics[width=\linewidth]{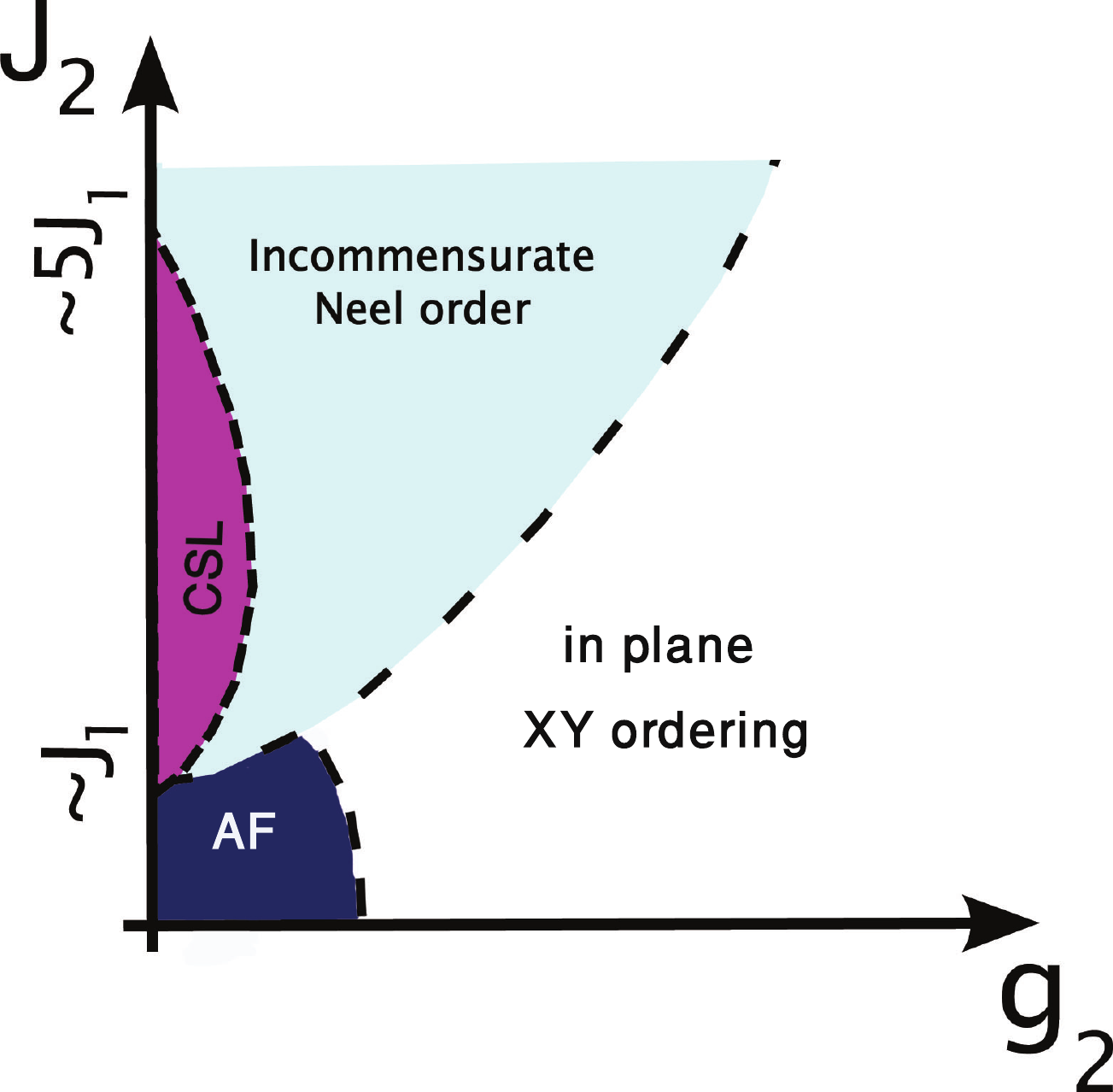}
\caption{(Color online) Our proposal for the phase diagram of the KMH model. CSL stands for chiral spin liquid phase and AF denotes out of plane Neel ordering. XY ordering refers to the Neel order along an axis in the plane, e.g. along the $x$ direction.}\label{Fig4}
\end{center}
\end{figure}

\subsection{Gauge theory of the topological gapped spin liquid phase}
In the topological gapped spin liquid phase, Schwinger fermions are gapped in the bulk and the only low energy excitation is the compact gauge field. Since spin up Schwinger fermions and spin down Schwinger fermions have opposite Hall conductances, the total Chern number vanishes as a result of the time reversal invariance of the ground state. Accordingly, the Chern-Simons action is absent in the low energy physics. Therefore, the gauge fluctuation is controlled by the gapless Maxwell term. However, instanton effect should also be taken into account due to the compact nature of the gauge symmetry.

In this phase, monopole insertion (instanton effect) adds a flux quantum of the gauge field to the system. Assuming $C_{\uparrow}=-C_{\downarrow}=+1$, a monopole insertion will increase the number of spin up Schwinger fermions by plus one and decreases the number of spin down by minus one. Therefore, the flux quantum of the gauge field carries a charge of the spin operator $S_z=1$ and flips the spin of Schwinger fermions. This means the density fluctuations of $S_z$ generates the magnetic flux of the gauge field. Since the gauge fluctuation is gapless and linearly dispersed, it can be viewed as the Goldstone mode of the system. Consequently, there should be a spontaneous XY ordering in the system with a nonzero $\braket{S^{+}_{i}}$ \cite{Wingho_2009_a}.

To conclude this section, at the mean field level we obtain a topological gapped spin liquid phase that does not break any lattice symmetry. The spectrum of fermions is gapped and so their fluctuations are suppressed. However, the gauge fluctuation is still gapless and should be taken into account. Among those fluctuations, the most relevant one are the instanton effects. Instantons proliferation condensates $S^{+}_{i}$ and $S^{-}_{i}$ operators and as a result there is an in-plane (XY) spin ordering in the groundstate. Instantons also gap out the gauge field. Therefore we finally end up with a state that is not a spin liquid and instead it breaks the spin rotational symmetry spontaneously.

\section{Summary and discussion}

We have studied the strongly correlated limit of the Kane-Mele-Hubbard model. We derived the Kane-Mele-Heisenberg interaction as the effective model for the spin degree of freedom. Using the Schwinger boson approach we obtained the phase boundary between the spin liquid and the magnetically ordered phases. The spin liquid phase happens to be a gapped state. Therefore, both charge and spin degrees of freedom are gapped in the spin liquid phase. To have a better insight of the spin liquid nature, we studied the Kane-Mele-Heisenberg model through Schwinger fermion approach. Within the mean field theory, we identified three types of spin liquids. The gauge theory of the Schwinger fermion model was discussed. We went beyond the mean field by taking the instanton effect into account. We discussed that instanton proliferation spoils all the spin liquid phases except the chiral spin liquid phase. However, it is not completely clear whether or not instantons will proliferate. Strong gauge fluctuations may generate pairing terms as well as hopping to farther neighbors. These terms can stabilize the spin liquid phase of matter. Moreover, large N limit studies of the compact SU(N) gauge fields yields the stability of the spin liquid phase. Further investigation of the fate of the spin liquids is needed to determine what happens to the proposed topological as well as the gapless spin liquid phases. Based on a combination of the Schwinger boson/fermion study, gauge theory and topological arguments we suggest that the phase diagram of the KMH is as depicted in Fig. [4].
\section{acknowledgement}
A.V. gratefully acknowledges useful discussions with Akbar Jafari and Dariush Heidarian.

%merlin.mbs 2010-03-15 4.21a (PWD, AO, DPC)

%\bibliography{Refs}
\end{document}